\documentclass[twocolumn,showpacs,preprintnumbers,amsmath,amssymb,aps,pra]{revtex4}

\usepackage{graphicx}
\usepackage{dcolumn}
\usepackage{bm}
\usepackage{color}

\begin{document}

\title{Light Transport in Random Media with ${\cal PT}$-symmetry}

\author{Samuel Kalish$^1$}
\author{Zin Lin$^1$}%
\author{Tsampikos Kottos$^{1,2}$}
\affiliation{$^1$Department of Physics, Wesleyan University, Middletown, Connecticut 06457}
\affiliation{$^2$Max-Planck-Institute for Dynamics and Self-Organization, 37073G\"ottingen, Germany}

\date{\today}

\begin{abstract}
The scattering properties of randomly layered optical media with ${\cal PT}$-symmetric index of refraction are studied 
using the transfer-matrix method. We find that the transmitance decays exponentially as a function of the 
system size, with an enhanced rate $\xi_{\gamma}(W)^{-1}=\xi_0(W)^{-1}+\xi_{\gamma} (0)^{-1}$, where $\xi_0(W)$ is the 
localization length of the equivalent passive random medium and $\xi_{\gamma}(0)$ is the attenuation/amplification length 
of the corresponding perfect system with a ${\cal PT}$-symmetric refraction index profile. While transmitance processes 
are reciprocal to left and right incident waves, the reflectance is enhanced from one side and is inversely suppressed 
from the other, thus allowing such ${\cal PT}$-symmetric random media to act as unidirectional coherent absorbers. 
\end{abstract}

\maketitle



{\it Introduction--}Wave propagation in naturally occurring or engineered complex media, is an interdisciplinary field of research 
that addresses systems as diverse as classical, quantum and atomic-matter waves. Despite this diversity, the wave 
nature of these systems provides a common framework for understanding their transport properties. One such 
characteristic is wave interference phenomena. Their existence results in 
a complete halt of wave propagation in random media which can be achieved by 
increasing the randomness of the medium. This phenomenon was predicted fifty years ago in the framework of quantum 
(electronic) waves by Anderson \cite{A58} and its existence has been confirmed in recent years in experiments with 
matter \cite{A08} and classical waves \cite{WBLR97,CSG00,HSPST08,C99,LAPSMCS08}. 

While the localization of classical waves has been well understood by now \cite{S90}, only in the last 
decade has light propagation in active random media been pursued intensively \cite{PK94,PACY10,YBCCT05,FPY94,JS99,
BPB96,PMB96,RD87}. Due to the absence of a conservation law for photons, light may be absorbed or amplified in the 
medium while phase coherence is preserved. This interplay of absorption or amplification and localization has been 
studied by using the Helmholtz equation with an imaginary dielectric constant of an appropriate sign. 
Several interesting results have been found, such as the dual symmetry of absorption and amplification for the 
average transmitance and the localization length \cite{BPB96,PMB96}, the sharpness of back scattering coherent 
peak and the statistics of super-reflectance and transmitance \cite{PK94,PACY10,YBCCT05,FPY94,JS99}. 

\begin{figure}[b]
\includegraphics[scale=0.33]{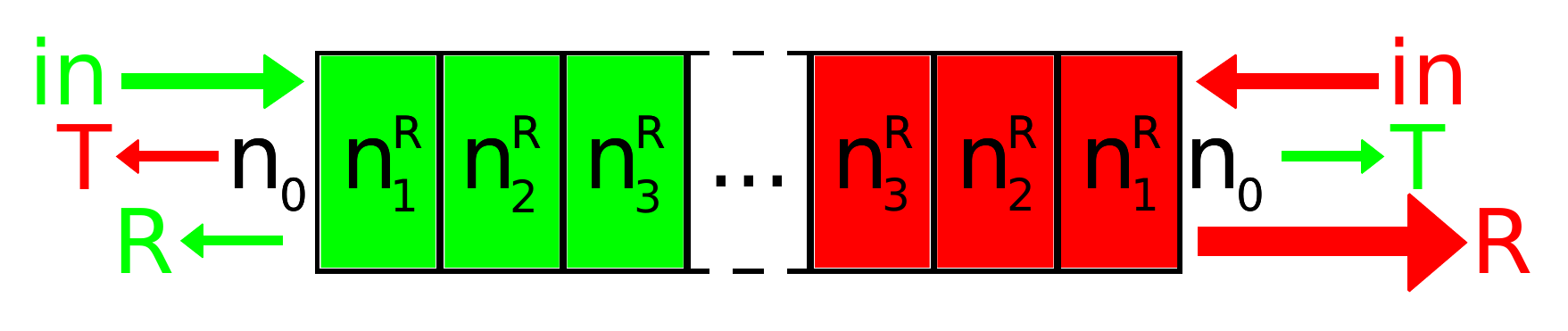}
\caption{(Color online) A one-dimensional ${\cal PT}$-symmetric multilayered random medium. The gain/loss refraction index 
profile is uniform (see Eq. (\ref{nzprofile})) with the loss side on the left (light color) and the gain side (dark color)
on the right of the structure. The real part of the refraction index contrast $n^{\rm R}$ is random, uniformly distributed
around $n_0$, and $n^{\rm R}(-z)=n^{\rm R}(z)$. For large enough system sizes 
(or strong disorder and/or large gain/loss) the system acts as a high performance absorber if the incident
wave is entering from the lossy side of the structure (light arrows), while it "super-reflects" if the incident wave
enters the structure from the gain side (dark arrows).}
\label{fig:draw} 
\end{figure}

Quite recently, the possibility of synthesizing a new family of artificial optical materials that instead rely on 
balanced gain and loss regions has been suggested \cite{MGCM08a,MMGC08b,RMGCSK10,K10,GSDMRASC09,RKGC10,ZCFK10,LRKCC11}. 
This class of optical structures deliberately exploits notions of parity (${\cal P}$)and time (${\cal T}$) symmetry 
\cite{BB98,BBDJ02,B07} as a means to attain altogether new functionalities and optical characteristics \cite{MGCM08a}. 
In optics, ${\cal PT}$ symmetry demands that the complex refractive index obeys the condition $n({\vec r})=n^*(-{\vec r})$, 
in other words the real part of the refractive index should be an even function of position, whereas the imaginary 
part must be odd. ${\cal PT}$ symmetries are not only novel mathematical curiosities. In a series of recent experimental 
papers ${\cal PT}$ dynamics have been investigated and key predictions have been confirmed and demonstrated \cite{RMGCSK10,
GSDMRASC09,SLZEK11,FAHXLCFS11}. These include among others, power oscillations \cite{MGCM08a,RMGCSK10,ZCFK10}, absorption 
enhanced transmition \cite{GSDMRASC09}, 
double refraction and non-reciprocity of light propagation \cite{MGCM08a}. In the nonlinear domain, such pseudo-Hermitian 
non-reciprocal effects can be used to realize a new generation of on-chip isolators and circulators \cite{RKGC10}. Other 
results within the framework of ${\cal PT}$-optics include the study of Bloch oscillations \cite{L09a}, the realization 
of coherent perfect laser absorbers \cite{L10b} and nonlinear switching structures \cite{SXK10}.

Work has also been done on disordered ${\cal PT}$ systems, with main focus on the spectral properties of the corresponding 
${\cal PT}$-Hamiltonians \cite{WKP10,SJ11}. However, relatively little has been done concerning their transport properties. 
In this paper, we will examine the 
transmitance and reflectance through one-dimensional (1D) ${\cal PT}$-symmetric systems with random index of refraction 
(see Fig. \ref{fig:draw}). We show that the exponential decay rate of transmitance, which defines the 
inverse localization length $\xi^{-1}$, is associated with the harmonic sum of the localization length $\xi_0(W)$ of a 
passive system with the same degree of randomness and the amplification length $\xi_{\gamma}(0)$ of a periodic 
${\cal PT}$-symmetric system with the same degree of gain/loss. Furthermore, we find that the asymptotic value of the reflectance
follows a single parameter scaling law which is dictated by the ratio $\Lambda=\xi_0(W)/\xi_{\gamma}(0)$.
Finally, we show that while the transmition processes are reciprocal to left and right incident waves, the reflection 
is enhanced from one side and is inversely suppressed from the other, thus allowing such ${\cal PT}$-symmetric random 
media to act as unidirectional coherent absorbers (see Fig. \ref{fig:draw}). 

{\it Mathematical Model--}We consider a one-dimensional ($1D$) active disorder sample having a random ${\cal PT}$-symmetric 
refractive index distribution $n(z)=n_{0} +n^{\rm R}(z) + i n^{\rm I}(z)$ in the interval $|z|<L/2$. The system is embedded 
in a homogeneous medium having a uniform refractive index $n_0$ for $|z|>L/2$ (see Fig. \ref{fig:draw}). Without loss of 
generality, below, we will assume that the refraction index $n_0$ outside the disordered medium is $n_0=1$. Here 
$n^{\rm R}$ represents the real index contrast and $n^{\rm I}$ the gain/loss spatial profile. In experimental realization
in optics \cite{MGCM08a,RMGCSK10}, these amplitudes are small, e.g. $n^{\rm R}, n^{\rm I}\ll n_0$. For simplicity, 
we will assume that the sample is composed of an even number, $L$, of layers of uniform width $d$, each with a 
constant real refractive index $n^{\rm R}(z_j-d/2\leq z\leq z_j+d/2)=n^{\rm R}_j$ given by a random variable with a uniform 
distribution between $(-W,+W)$ which satisfies the ${\cal PT}$-symmetric constraint $n^{\rm R}(z_j)=n^{\rm R}(-z_j)$.
Specifically we will assume 
\begin{eqnarray}
n(z)&=&n_0+n^{\rm R}(z)+i\gamma \text{ for } -L/2<z<0 \notag \\
    &=&n_0+n^{\rm R}(z)-i\gamma \text{ for } 0<z<L/2 
\label{nzprofile}
\end{eqnarray}
where $\gamma\geq 0$ is a fixed gain/loss parameter $n^{\rm I}$.
Although the majority of our simulations below have been done for $n^{\rm I}(\pm z)=\mp \gamma$, we have also checked 
that our results apply for the case that  $n^{\rm I}(\pm z)=\mp\gamma +\delta n^I$ where $\delta n^I$ is a random 
variable given by a uniform distribution centered at zero. Since the qualitative features remain the same, we will
not distinguish between these two cases. 
In this arrangement, a time-harmonic electric 
field of frequency $\omega$ obeys the Helmholtz equation:
\begin{equation}
\label{Helmholtz}
{\partial^2 E (z)\over \partial z^2} + {\omega^2 \over c^2} n^2(z) E(z) = 0\,\,\,.
\end{equation}
Eq. (\ref{Helmholtz}) admits the solutions 
$E_0^{-}(z)=E_{f}^- e^{ikz} + E_{b}^- e^{-ikz}$ for $z< -L/2$ and $E_0^{+}(z)=E_{f}^+ e^{ikz} + E_{b}^+ e^{-ikz}$, for $z> L/2$ 
where the wave-vector $k= n_0\omega/c$. 
The amplitudes of forward and backward propagating waves outside the disorder domain are related via
the transfer matrix $M$:
\begin{eqnarray}
\label{transfer}
\left(\begin{array}{c}
E_{f}^+\\
E_{b}^+
\end{array}\right) & = &
\left(\begin{array}{cc}
M_{11}&M_{12}\\
M_{21}&M_{22}
\end{array}\right)
\left(\begin{array}{c}
E_{f}^-\\
E_{b}^-
\end{array}\right) 
\end{eqnarray}
The transmission and reflection amplitudes for left (L) and right (R) 
incidence waves, can be obtained from the boundary conditions $E_b^+=0$ ($E_f^-=0$) respectively, and are defined 
as $t_L\equiv {E_f^{+}\over E_f^{-}}$, $r_L\equiv {E_b^{-}\over E_f^{-}}$; ($t_R\equiv {E_b^{-}\over E_b^{+}}$; 
$r_R\equiv {E_f^{+}\over E_b^{+}}$). These can be expressed in terms of the transfer matrix elements as 
$t_L=t_R=t={1\over M_{22}}; r_L= -{M_{21}\over M_{22}}; r_R = {M_{12}\over M_{22}}$ \cite{CDV07,M09}.
While the transmitance $T=|t|^2$ for left or right incidence is the same, this is not necessarily the case for the
left and right reflectance $R_L=|r_L|^2$ and $R_R=|r_R|^2$ respectively. Furthermore, from the above relations one 
can deduce that $r_{L}r_{R}^*=(1-|t|^2)$ \cite{L10b,Stone}. Thus one establishes the conservation relation \cite{L10b,Stone}:
\begin{equation}
 \sqrt{R_{L}R_{R}}=|1-T| \label{eqn3}
\end{equation}
When considering Hermitian scattering systems, where $R_{L}=R_{R}$ and $T\le1$, this relation recovers the 
familiar $T+R=1$. The non-Hermitian systems studied in this paper, however, exhibit non-identical reflectances, 
and Eq.~(\ref{eqn3}) describes the connection between the two.

Below we investigate the scaling properties of transmitance $T$ and left and right reflectances $R_L,R_R$ from such 
optical structures, with respect to the disorder strength $W$, the gain/loss parameter $\gamma$, and the wave-vector 
of the incoming wave. We have used various random refraction index contrasts $W\in 
(0,0.5)$, and gain/loss parameters $\gamma \in (0,0.01)$ 
which is typical for optical media. We use slabs with $L=10$ to $L=10^4$ number of layers, each having width $d=1$. The 
logarithmic averages $\langle\ln T\rangle$ and $\langle\ln R\rangle$ are performed over $10^{4}$ disorder 
realizations.

\begin{figure}
\includegraphics[scale=0.33]{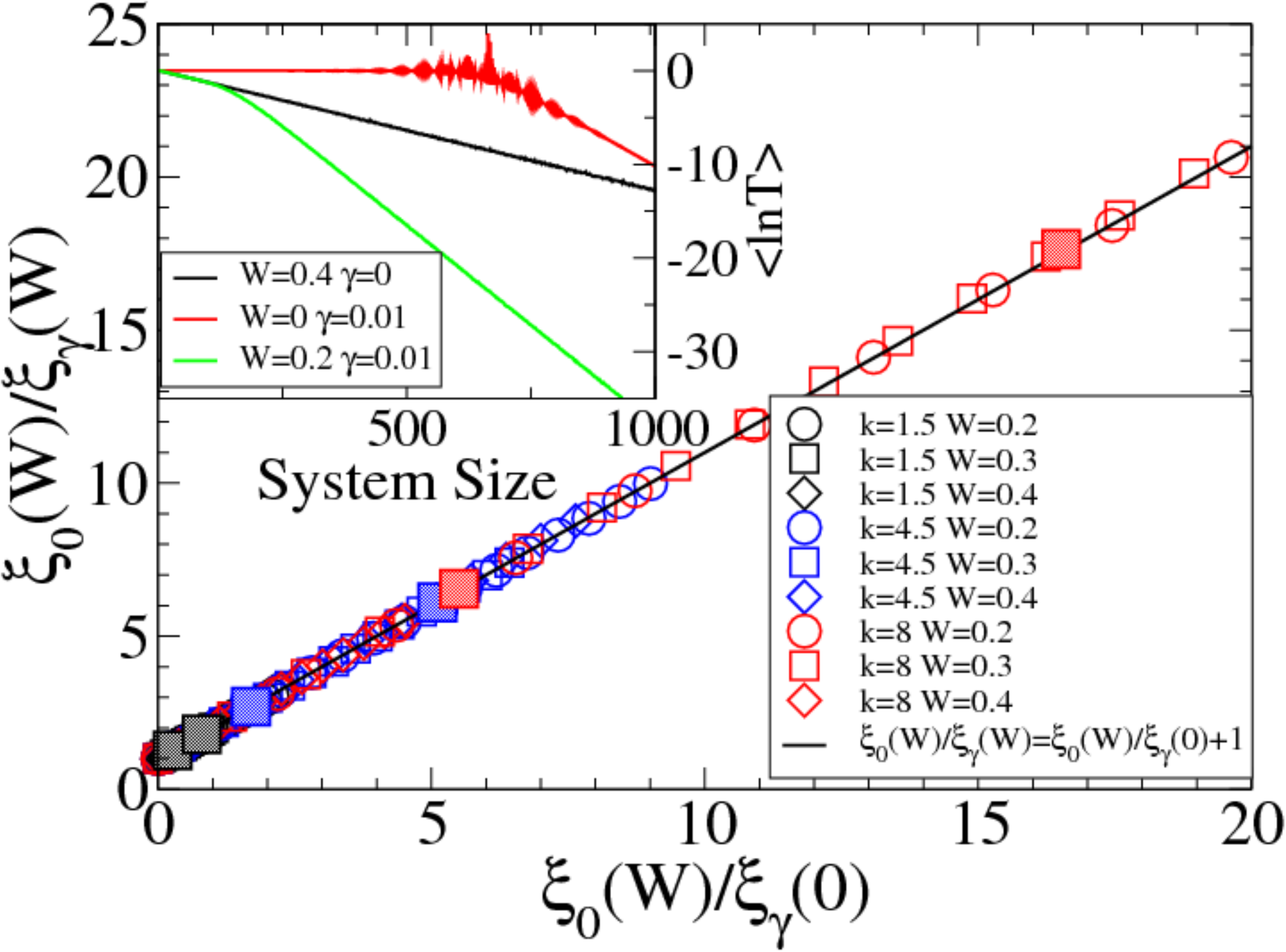}
\caption{
(Color online) The numerically extracted localization length $\xi_{\gamma}(W)$ for various gain/loss parameter $\gamma$ (not 
indicated in the figure) is plotted, rescaled with $\xi_0(W)$ versus the scaling parameter $\xi_0(W)/\xi_{\gamma}
(0)$. The symbols and colors indicate different  wavelengths $k$ of the incident wave, and disorder strengths $W$. 
The black line indicates the theoretical prediction of Eq.~( \ref{eqn1}). The meshed symbols correspond to some typical 
$\xi_{\gamma}(W)$, for the scenario where the imaginary part of the refraction index is $n^{I}=\gamma+\delta n^{I}$ where 
$\delta n^{I}$ is a random variable uniformly distributed around zero. In the inset we report $\langle \ln T\rangle$ 
against the system size for: a system with only gain/loss and $n^R=n_0$ constant (upper red); a random layer medium with 
$n^{\rm I}=0$ (middle black); and a ${\cal PT}$-symmetric disorder medium (lower green).
}
\label{fig:LocLengthScaledInsert} 
\end{figure}

{\it Transmittance--}The transport properties of passive (no gain or loss) disorder systems have been thoroughly studied. At 
large length-scale, such systems exhibit an exponential decay in transmittance (see middle black line in the inset of Fig.~
\ref{fig:LocLengthScaledInsert}). The associated inverse decay rate $\xi_0(W)$, reflects the degree of randomness and 
it is defined as 
\begin{equation} 
1/\xi_{0}(W)\equiv -\lim_{L\rightarrow \infty} \langle \ln T \rangle/L
\end{equation}
For 1D random media we have that $\xi_{0}(W) \sim 1/W^{2}$ \cite{S90}.

The other limiting case of an ordered ${\cal PT}$-symmetric medium, can also be treated analytically. One can explicitly 
solve for the electric field inside the perfect ${\cal PT}$ layered structure subject to scattering boundary conditions. 
The resulting expression for the transmitance reads:
\begin{widetext}
\begin{eqnarray}
 T={8 \left(1+\gamma ^2\right)^2
\over 
\gamma ^2 \left(4+\gamma ^2\right) \Big(\gamma^2\cosh(2kL\gamma)- \cos(2kL)\Big) 
+ 8+5 \gamma ^4-\gamma ^6
+ 4 \gamma ^2 \Big(1+\cos(kL) \cosh(kL\gamma) - \gamma \left(2+\gamma ^2\right) \sin(kL) \sinh(kL\gamma)\Big)}
\label{Tperfect}
\end{eqnarray}
\end{widetext}
For large $L$-values, the term involving $\cosh(2kL\gamma)$ becomes dominant and the transmitance decays exponentially as 
shown in the inset of Fig.~\ref{fig:LocLengthScaledInsert} (see upper red line). For the experimentally relevant case $\gamma \ll 1$, 
the asymptotic decay in transmitance can be found from Eq. (\ref{Tperfect}) to be
\begin{eqnarray}
\label{tperf_inf}
T_{\infty} \approx \frac{16 e^{-2k\gamma L}}{\gamma^4 (4+\gamma^2)}
\end{eqnarray}
The corresponding decay rate of transmitance is
\begin{equation}
\label{Tperfasym}
{1\over \xi_\gamma(0)}\equiv -\lim_{L\rightarrow \infty} {1\over L} \ln T \rightarrow 2 k \gamma
\end{equation}
which can serve as an operative definition of the so-called attenuation/amplification length $\xi_\gamma(0)$.

On the other hand, for system sizes $L$ smaller than a critical length-scale $L_c$ the transmitance remains approximately 
constant $T\approx 1$. Near the critical length $L\approx L_c$, large oscillations in the transmitance emerge (see upper red line 
in the inset of Fig.\ref{fig:LocLengthScaledInsert}) after which the transmitance decays according to the expression given by Eq. (\ref{tperf_inf}).
The value of $L_c$ can be evaluated approximately by the condition $T_{\infty}(L=L_c)=1$ which leads to the following 
expression 
\begin{equation}
\label{Lc}
L_c\approx \frac{1}{2k\gamma}\ln \Big(\frac{16}{\gamma^4 (4+\gamma^2)}\Big). 
\end{equation}
The existence of a critical length-scale $L_c$ is characteristic of gain media and is associated with the lasing threshold
for which $T$ diverges. Below this length stimulated emission enhances transmitance through the gain medium. On larger length
scales stimulated emission reduces transmitance. The slopes of $\ln T$ at both sides of the maximum are approximately symmetric.
In contrast, in the case of a ${\cal PT}$-symmetric refraction index the increase of the transmitance for $L<L_c$ which is due
to the gain, is balanced by the equal amount of loss which is symmetrically arranged inside the medium. As a result $T\approx 1$ 
for $L<L_c$. Nevertheless, this gain/loss balance, is not able to smooth out the diverging behavior of $T$ near the lasing 
threshold (see inset of Fig.~\ref{fig:LocLengthScaledInsert}).

Let us finally consider the case of ${\cal PT}$-symmetric disordered slab geometry. A representative behavior of the 
transmitance $T$ as a function of the system size $L$ is shown in the inset of Fig.~\ref{fig:LocLengthScaledInsert} (see lower green line). To understand 
the exponential decay of $T$, one needs to consider the simplified geometry with index of refraction given by Eq. (\ref{nzprofile}).
For the case of Eq. (\ref{nzprofile}) the transfer matrix of the total ${\cal PT}$-symmetric system is 
the product of the transfer matrix $M_{l}$ associated with the lossy sub-system and $M_{g}$ associated with the 
gain sub-system. The corresponding transmitance through the combined system is given by
\begin{equation}
\label{T_PTD}
 T=\frac{|T_{l}T_{g}|}{|1-r_{l}r_{g}|^2}
\end{equation}
It is thus sufficient to know the scaling behavior of each of the terms on the rhs of Eq. (\ref{T_PTD}) in order to predict 
the scaling behavior of $T$. These terms have been studied in Refs. \cite{PMB96}, where it was found that both absorption
and amplification lead to the same exponential decay of the transmitance which in both cases is enhanced with respect to a 
passive disordered medium by the strength of the gain (or loss) rate i.e.  $\langle \ln T_{l,g} \rangle=-(2k\gamma+
\xi_{0}(W)^{-1})L/2$.
Somewhat 
counter-intuitively, the sample with amplification also exhibits exponentially decaying transmitance due to the enhanced internal 
reflections from the boundaries. Using the duality relation \cite{PMB96} $r_{l}r_{g}^{*}=1$ for the reflection 
of an amplifying or attenuating medium (with the same rate of gain or loss respectively) applied for $L/2>\xi_{0}(W)$ we get
\begin{eqnarray}
\label{Tgl}
\langle \ln T \rangle&=&\langle \ln T_{l} \rangle+\langle \ln T_{g}\rangle 
         -2\langle \ln|1-r_{l}r_{g}| \rangle \\
       &=&\langle \ln T_{l}\rangle+\langle \ln T_{g}\rangle 
       -2\langle\ln|2(1-\cos(2\theta))|\rangle\notag 
\end{eqnarray}
where $\theta$ is the phase of the reflection amplitude $r_l$. Assuming that $\theta$ is a random variable uniformly 
distributed on the interval $[0,2\pi]$ \cite{RD87}, we get that the last term after performing the
average over the random variable $\theta$ is finite. Therefore we get:
\begin{equation}
\lim_{L\rightarrow \infty} \frac{\langle \ln T\rangle}{L} = 
       =-(2k\gamma+\xi_{0}(W)^{-1})
\end{equation}
From the above argument, we conclude that the localization length for a ${\cal PT}$-symmetric disorder medium is:
\begin{equation}
\xi_{\gamma}(W)^{-1}=\xi_{\gamma}(0)^{-1}+\xi_{0}(W)^{-1}
\label{eqn1}
\end{equation}
In Fig.~\ref{fig:LocLengthScaledInsert}, we show the results of our numerical simulations for a disordered ${\cal PT}$-symmetric
sample. The extracted localization length nicely follows the scaling behavior indicated by our theoretical arguments.

\begin{figure}[b]
\includegraphics[scale=0.33]{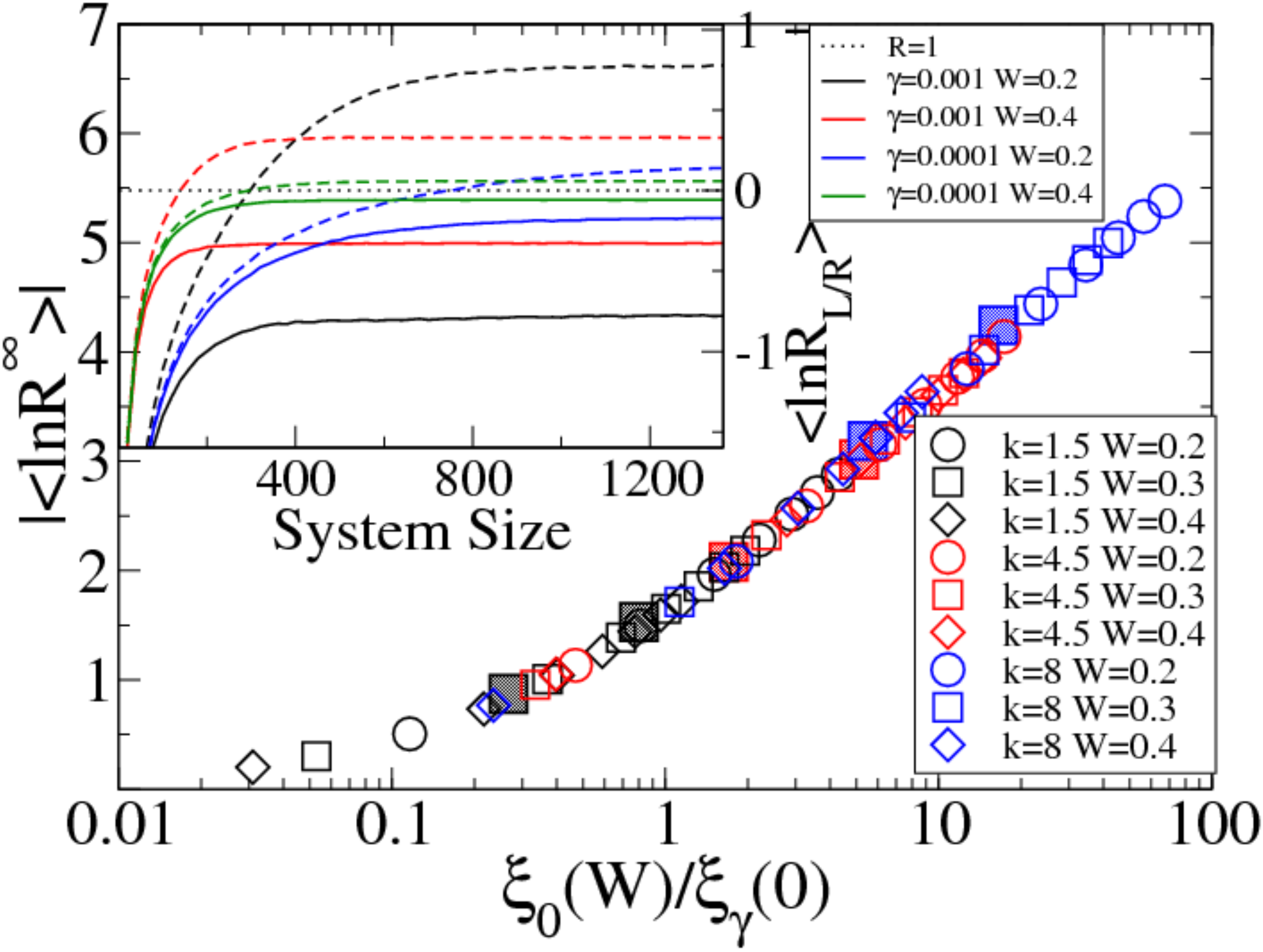}
\caption{(Color online)
Insert: Typical reflectances $R_{\rm L,R}$ versus the system size (number of layers $L$). The main figure
displays the asymptotic value of $|\log R^{\infty}|$ versus the scaled parameter $\Lambda$. For large values of $\Lambda$ the
$|\log R^{\infty}|$ increases indicating that $R_{\rm L}^{\infty}$ (i.e. reflectance for an incident wave entering the 
sample from the lossy side) diminishes. In this domain, the sample acts as a unidirectional absorber. We use the same symbol 
and color coding for our data as in Fig. \ref{fig:LocLengthScaledInsert}.
}
\label{fig:ReflectionScaled} 
\end{figure}

{\it Reflectance--}
We proceed with the analysis of the reflectances. In the case of random media with only gain or loss it was 
found in Ref. \cite{PMB96} that the reflectances of two {\it distinct} disordered systems, one with gain
strength $-\gamma$ and the other with loss strength $\gamma$ satisfy the following reciprocal relation between them:
\begin{equation}
 R_{gain}R_{loss}=1 
\label{eqn2}
\end{equation}
Specifically it was found that for gain media $R_{\rm gain}>1$, while for lossy media we have via Eq. (\ref{eqn2}) the 
reciprocal behavior $R_{\rm loss}=R_{\rm gain}^{-1}<1$. It is important to stress here that such systems do not distinguish 
between left and right incidence, that is $R_{L}=R_{R}$ for each of the cases.

On the other hand, ${\cal PT}$-symmetric systems distinguish the reflection between left or right incident wave, that is, 
$R_{L}\neq R_{R}$ in general (see previous discussion). This phenomenon has already been observed for periodic 
${\cal PT}$-symmetric structures in Ref.~\cite{LRKCC11}. Moreover, in the presence of random index of refraction, we have 
previously concluded that the transmitance is effectively diminished exponentially with a rate $1/
\xi_{\gamma}(W)$ given by Eq. (\ref{eqn1}). Using the conservation relation Eq.~(\ref{eqn3}) we get
\begin{equation}
\label{RL_RR}
R_{\rm R}R_{\rm L} \rightarrow 1. 
\end{equation}
Although this relation is similar to Eq. (\ref{eqn2}), it should be emphasized once more that in the case of 
${\cal PT}$-symmetric disorder media the medium behaves simultaneously as a gain medium (i.e. having $R_L>1$) and as a lossy medium 
(i.e. it can enhance absorption of incoming coherent waves $R_R<1$). The reciprocity of the left and right reflectances is 
clearly demonstrated for some representative cases in the inset of Fig.~\ref{fig:ReflectionScaled}. 

A natural question is associated with the scaling behavior of the asymptotic value of
the reflectances $R_{L,R}^{\infty}$ as a function of the disorder strength $W$ and the gain/loss parameter $\gamma$. We 
speculate that a one-parameter scaling law describes the asymptotic reflectance i.e.
\begin{equation}
\label{Rscaling}
R_{L,R}^{\infty}(\gamma,W)=f(\Lambda), \quad {\rm where}\quad \Lambda=\xi_0(W)/\xi_{\gamma}(0)
\end{equation}
Here, $R_{L,R}^{\infty}$ is the geometric mean of asymptotic reflectance, that is, $R_{L,R}^{\infty}=\exp(\langle \ln R_{L,R}
^\infty \rangle)$. We have tested our hypothesis numerically. To this end, we have extracted $R_{L,R}^{\infty}$ from our data 
for various values of $\gamma$ and $W$ and plot them against the scaling variable $\Lambda$. The results are presented in 
the main part of Fig.~\ref{fig:ReflectionScaled}. We find that for realistic values of the gain/loss parameter $\gamma\leq 
10^{-2}-10^{-3}$ the data nicely follow the one-parameter scaling hypothesis (\ref{Rscaling}). As the scaling parameter 
$\Lambda$ increases (either by decreasing $W$ or by increasing the gain/loss parameter 
$\gamma$), the asymptotic value $R_{L,R}^{\infty}$ decreases/increases. Such a behavior allows us to use the proposed 
structure as a unidirectional coherent absorber that can increase absorption by tuning up the scaling parameter $\Lambda$.
We want to mark that our structure, is different from the one suggested in Ref. \cite{CS11}, where it is 
shown that a disordered system with a single absorbing element causes coherent enhanced absorption if the 
phases of the input field are appropriately manipulated. Instead, we 
are addressing a different problem, where we have broadband absorption in one direction without a need for phase manipulation
of the incoming wave.

{\it Conclusions--} 
We have investigated the transport properties of one-dimensional ${\cal PT}$-symmetric disordered layers. We have found that
the localization length $\xi_{\gamma}(W)$, defined as the inverse decay rate of the transmitance, is smaller than the localization
length of the passive disordered system $\xi_0(W)$ and from the absorption/amplification length $\xi_{\gamma}(0)$ of a periodic 
${\cal PT}$-symmetric medium. At the same
time the reflectance depends on the direction of the incident wave: while for incident waves entering the medium from the gain side 
it is enhanced, it is suppressed if the wave enters the medium from the lossy side. The reduction/enhancement of the reflectance
is dictated by a one parameter scaling $\Lambda=\xi_0(W)/\xi_{\gamma}(0)$ and allows us to use such structures as unidirectional
quasi-perfect coherent absorbers.

{\it Acknowledgments :} We acknowledge support by an AFOSR No. FA 9550-10-1-0433 grant and by an NSF ECCS-1128571 grant.


\end{document}